\documentclass[sigconf]{acmart}

\renewcommand\footnotetextcopyrightpermission[1]{}
\usepackage{adjustbox}
\usepackage{makecell}
\usepackage{multicol}
\usepackage{multirow}
\usepackage{hhline}
\usepackage{booktabs}
\usepackage{color}
\usepackage{xcolor}
\usepackage{amsmath}
\usepackage{CJK}
\usepackage{tabularx}

\title[]{Learning Shared Sentiment Prototypes for Adaptive\\ Multimodal Sentiment Analysis}

\author{Chen Su}
\affiliation{%
  \institution{University of Science and Technology of China}
  \city{}
  \country{}
}
\email{suchen4565@mail.ustc.edu.cn}

\author{Yuanhe Tian}
\affiliation{%
  \institution{Zhongguancun Academy}
  \city{}
  \country{}
}
\email{yhtian94@gmail.com}

\author{Yan Song$^*$}
\affiliation{%
  \institution{University of Science and Technology of China}
  \city{}
  \country{}
}
\email{clksong@gmail.com}

\renewcommand{\shortauthors}{}

\begin{document}

\begin{abstract}
Multimodal sentiment analysis (MSA) aims to predict human sentiment from textual, acoustic, and visual information in videos. Recent studies improve multimodal fusion by modeling modality interaction and assigning different modality weights. However, they usually compress diverse sentiment cues into a single compact representation before sentiment reasoning. This early aggregation makes it difficult to preserve the internal structure of sentiment evidence, where different cues may complement, conflict with, or differ in reliability from each other. In addition, modality importance is often determined only once during fusion, so later reasoning cannot further adjust modality contributions. To address these issues, we propose PRISM, a framework that unifies structured affective extraction and adaptive modality evaluation. PRISM organizes multimodal evidence in a shared prototype space, which supports structured cross-modal comparison and adaptive fusion. It further applies dynamic modality reweighting during reasoning, allowing modality contributions to be continuously refined as semantic interactions become deeper. Experiments on three benchmark datasets show that PRISM outperforms representative baselines.\footnote{The code is available at \url{https://github.com/synlp/PRISM}.}
\end{abstract}

\maketitle

\renewcommand{\thefootnote}{\fnsymbol{footnote}}
\footnotetext[1]{Corresponding author.}

\renewcommand{\thefootnote}{\arabic{footnote}}

\section{Introduction}

\begin{figure*}
    \centering
    \includegraphics[width=1\linewidth, trim=0 15 0 0]{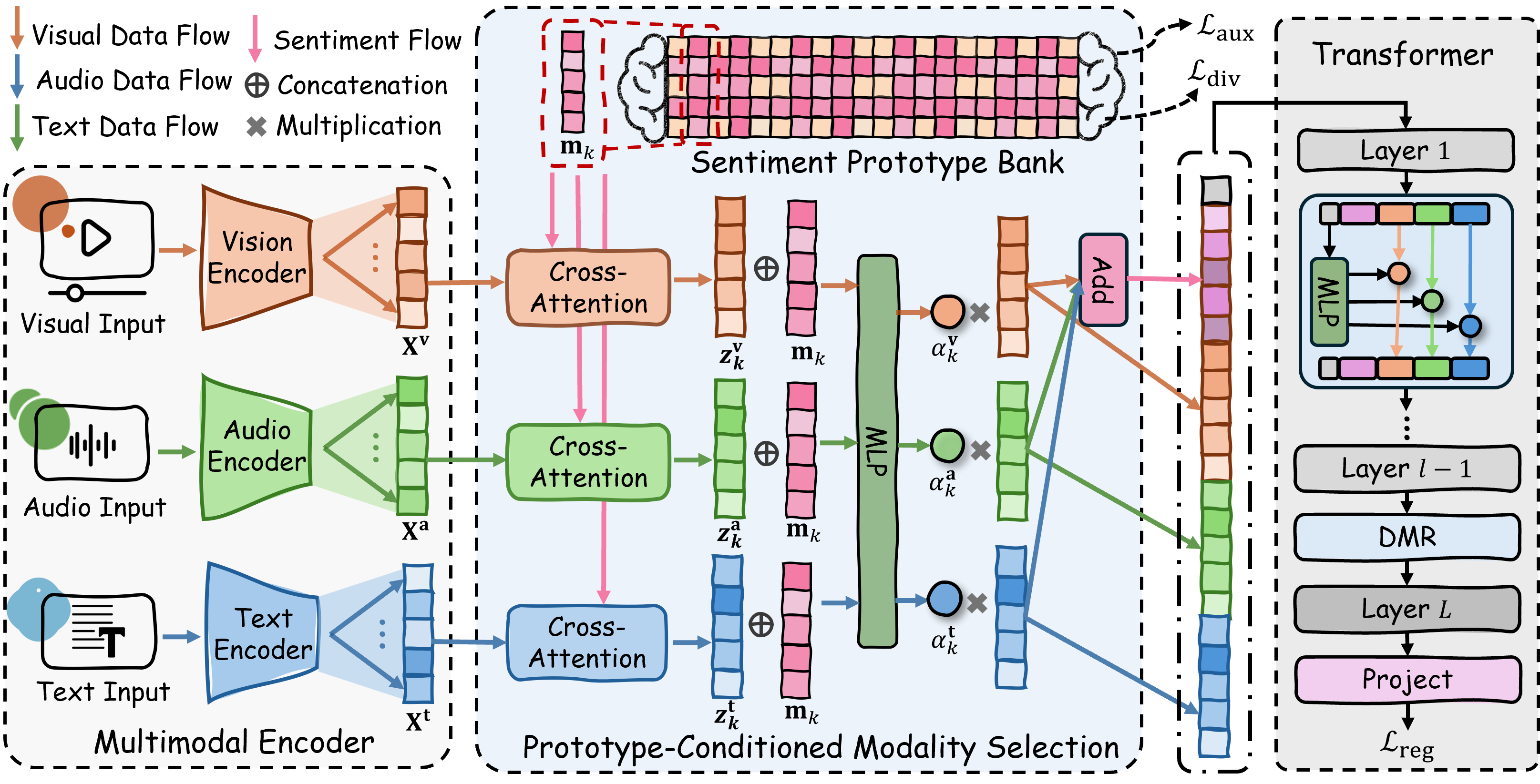}
    \caption{Overview of the PRISM framework. The left panel illustrates the multimodal encoding stage, where visual, acoustic, and textual inputs are independently encoded into modality-specific feature sequences. The middle panel presents the sentiment prototype bank (SPB) and prototype-conditioned modality selection where shared sentiment prototypes interact with each modality through cross-attention to extract prototype-aligned unimodal responses, which are then adaptively combined by predicting prototype-wise modality weights to form the fused prototype sequence. The right panel shows the Transformer backbone with dynamic modality reweighting (DMR), where modality contributions are continuously revised during reasoning before the final projection produces the sentiment prediction.}
    \label{fig:model}
    \vspace{-0.3cm}
\end{figure*}

As a fundamental task in multimodal understanding, multimodal sentiment analysis (MSA) aims to predict human sentiment from videos by jointly modeling textual, acoustic, and visual information \cite{zeng2007survey, d2015review, poria2017review, tsai2019multimodal, geetha2024multimodal}.
Since human sentiment is often expressed through words, vocal characteristics, and facial behaviors simultaneously, relying on a single modality is often insufficient for obtaining complete sentiment cues \cite{baltruvsaitis2018multimodal, liu2022make, guan2024hallusionbench}.
By integrating multimodal information, MSA enables more accurate understanding of speaker sentiment and thus holds substantial value for affective computing \cite{picard1997affective, cambria2017affective, poria2017review}, human--computer interaction \cite{pantic2011social, devillers2021human}, and content understanding \cite{melville2009sentiment, doctor2016intelligent, jiang2020snapshot, ezzameli2023emotion, yang2023context}.

Realizing this goal requires effectively fusing information from heterogeneous modalities, and existing research has explored this challenge through progressively evolving fusion strategies \cite{schuller2013paralinguistics, poria2017review, zadeh2018multi, guo2022dynamically}.
Early representative approaches \cite{zadeh2017tensor, liu2018efficient, zadeh2018memory, tsai2019multimodal} model inter-modal interactions through tensor composition or cross-modal attention and usually treat all modalities symmetrically.
As textual content often conveys sentiment more explicitly than acoustic and visual signals, later studies increasingly adopt text-guided designs that use language to refine or filter non-textual information \cite{rahman2020integrating, wu2021text, zhang2023almt}.
More recent approaches further explore adaptive modality weighting \cite{wang2020transmodality, yang2020cmbert} and compressed fusion \cite{han2021mmim, wu2023dbm}, using gating \cite{huang2024simsuf}, bottleneck compression \cite{wen2025dashfusion}, or dynamic attention \cite{feng2024kuda, zhou2025dpdf} to adjust modality contributions and reduce redundancy.
%
%
Despite these advances, existing approaches still overlook a basic property of multimodal sentiment evidence.
Within each modality, sentiment evidence is not homogeneous, but consists of multiple affective cues with internal structure.
For example, a single utterance may simultaneously contain lexical polarity, emphasis, hesitation, facial tension, or expression--tone inconsistency, and these cues do not contribute equally to sentiment understanding.
However, most existing methods still compress each modality, or the fused multimodal evidence, into a single compact representation before sentiment reasoning \cite{yang2023confede, jiang2025ddse, wen2025dashfusion}.
Such early aggregation blends heterogeneous affective cues into an undifferentiated vector, making it difficult to preserve which cues are complementary, which are conflicting, and which are more reliable for the current sample.
Consequently, subsequent reasoning operates on a representation in which fine-grained affective distinctions are partially collapsed, limiting the model’s ability to recover their respective roles afterward.
As a result, subtle but important sentiment signals may be weakened or lost during fusion.
A further issue is that modality importance is usually determined only once at the fusion stage.
Even when existing approaches adaptively weight modalities for each sample, this adjustment is typically completed before subsequent reasoning begins \cite{yang2020cmbert, wang2026multimodal}.
Once the fused representation is formed, the model lacks a mechanism to continuously suppress misleading modality evidence or strengthen especially informative cues during later reasoning.
This is restrictive because modality reliability is not only sample-dependent, but may also become clearer as higher-level semantic interactions are progressively formed during reasoning.

To address these limitations, we propose PRISM (\textbf{P}rototype \textbf{R}easoning with \textbf{I}ntegrative \textbf{S}entiment \textbf{M}odeling), a framework for multimodal sentiment analysis that explicitly organizes multimodal evidence into structured affective components and preserves modality-level controllability throughout reasoning.
Rather than directly collapsing each modality into a single summary vector, PRISM first decomposes multimodal evidence into a small set of shared sentiment prototypes that serve as compact affective reference points.
Each prototype is intended to capture one recurrent type of sentiment-relevant evidence, so that the final representation is formed as an organized set of prototype-level responses instead of an undifferentiated global mixture.
Specifically, PRISM employs a shared sentiment prototype bank composed of learnable prototypes, where each prototype queries every modality sequence through cross-attention.
Because the same prototypes are applied to all modalities, the responses from text, audio, and visual streams are organized into aligned prototype slots.
This shared slot structure gives PRISM two important properties.
First, it imposes an explicit organization on multimodal evidence, making different affective cues more separable and directly comparable across modalities at the same prototype position.
Second, it provides a common reference basis for assessing modality reliability, because the model no longer evaluates a modality from an isolated global summary, but from how that modality responds to the same set of affective queries as the other modalities.
Based on this slot-wise organization, PRISM further estimates modality reliability by using each modality-specific prototype response together with the corresponding shared prototype, and then performs adaptively weighted fusion.
In this way, structured affective extraction and modality evaluation are unified within the same mechanism.
The prototypes not only determine how evidence is decomposed, but also define the basis on which modality contributions are judged.
To preserve modality-level controllability after fusion, PRISM does not discard the original modality-specific evidence once fusion is completed.
Instead, the fused prototype tokens and the original modality-specific tokens are jointly processed by a Transformer \cite{vaswani2017attention} backbone, where layer-wise gates continuously regulate modality contributions during reasoning.
This design allows the model to refine its reliance on different modalities as semantic interactions become deeper, rather than fixing modality importance only once at the fusion stage.
Experiments on three benchmark datasets, including CMU-MOSI, CMU-MOSEI, and CH-SIMS, show that PRISM outperforms representative baselines.
Ablation studies and further analyses also verify the contribution of each component.
\section{Related Work}
\label{sec:related}

\subsection{Multimodal Fusion for Sentiment Analysis}
\label{sec:rw_fusion}

Multimodal fusion is a central problem in MSA.
Early approaches model explicit inter-modal interactions through feature composition.
TFN \cite{zadeh2017tensor} uses tensor outer products to capture combinatorial interactions, LMF \cite{liu2018efficient} reduces this cost through low-rank decomposition, and MFN \cite{zadeh2018memory} tracks cross-view dynamics with a memory mechanism.
With the rise of Transformer architectures, attention-based fusion becomes dominant.
MulT \cite{tsai2019multimodal} uses directional cross-modal attention to handle unaligned sequences, while MAG-BERT \cite{rahman2020integrating} injects acoustic and visual information into pretrained language representations through gated adaptation.
Another line of work focuses on representation learning before fusion.
MISA \cite{hazarika2020misa} decomposes each modality into modality-invariant and modality-specific subspaces, and ConFEDE \cite{yang2023confede} further improves this decomposition through text-anchored contrastive learning.
As text representations are usually more informative than acoustic and visual features extracted by conventional tools, later approaches increasingly adopt text-guided designs.
CENet~\cite{wang2023cenet} enhances text representations with asynchronous nonverbal cues, ALMT~\cite{zhang2023almt} uses hierarchical language features to filter noisy nonverbal information and build a complementary hyper-modality representation, and DDSE~\cite{jiang2025ddse} further combines feature decoupling with text-centric progressive interaction by separating each modality into public and private components and using a Mamba to preserve key textual sentiment cues during fusion.
Recent work further studies dynamic weighting and compressed fusion.
KuDA~\cite{feng2024kuda} performs per-sample modality weighting from unimodal sentiment clues.
MMIM~\cite{han2021mmim} and DBF~\cite{wu2023dbm} improve fusion through mutual-information-guided compression, while DashFusion~\cite{wen2025dashfusion} applies hierarchical bottleneck fusion to progressively compress multimodal information.
DPDF-LQ~\cite{zhou2025dpdf} combines a learnable-query global path with a local path to balance global and fine-grained sentiment cues.
Despite these advances, existing approaches still overlook that each modality contains multiple affective cues with internal structure.
Most approaches fuse multimodal evidence in an aggregated manner, without explicitly preserving such structured affective information, so subtle but important sentiment cues are weakened during fusion.
In contrast, PRISM decomposes multimodal evidence into a small set of sentiment prototypes and further uses them to estimate modality reliability.
This design aligns and fuses sentiment cues in a shared prototype space, rather than directly compressing heterogeneous evidence into a single undifferentiated representation.

\subsection{Prototype Learning and Structured Representations}
\label{sec:rw_prototype}

Using a small set of representative vectors to capture the structure of high-dimensional inputs is a recurring idea in machine learning.
Early prototype-based methods in metric learning, such as Matching Networks and Prototypical Networks, show that representative vectors can support structured comparison in a learned space \cite{vinyals2016matching,snell2017prototypical}.
Subsequent work further shows that making such vectors learnable enables more flexible structured extraction.
For example, \citet{lin2017structured} use multiple learnable attention vectors to extract different semantic aspects from a sentence, and NetVLAD performs trainable feature aggregation through learnable cluster centers \cite{arandjelovic2016netvlad}.
Transformer architectures extend this idea into a learnable-query paradigm.
Set Transformer and Perceiver use a fixed number of learnable latent vectors as attention bottlenecks to compress variable-sized inputs \cite{lee2019set,jaegle2021perceiver,jaegle2022perceiverio}.
Slot Attention and DETR use learnable slots or object queries to extract structured representations through cross-attention \cite{locatello2020slot,carion2020detr}.
A similar query bottleneck design is also widely used in vision-language models, where Flamingo, BLIP-2, and InstructBLIP use learnable queries to compress visual inputs into a fixed set of tokens for downstream language modeling \cite{alayrac2022flamingo,li2023blip2,dai2023instructblip}.
However, these methods mainly use representative vectors for general-purpose compression, rather than multimodal sentiment analysis.
They also do not enforce a shared slot basis for comparing sentiment evidence across modalities.
In contrast, PRISM applies the same shared prototype bank to text, audio, and visual modalities, so prototype responses are aligned by slot and directly comparable across modalities.
This shared structure enables the prototypes to support both structured affective extraction and modality evaluation within a unified mechanism.
\section{The Approach}
As shown in Fig. \ref{fig:model}, given multimodal input $\mathcal{X}=\{\mathbf{X}^{\mathrm{v}},\mathbf{X}^{\mathrm{a}},\mathbf{X}^{\mathrm{t}}\}$, the PRISM framework sequentially performs modality encoding, prototype-driven extraction based on sentiment prototype bank (SPB), prototype-conditioned selection (PCS), and dynamic modality-reweighted reasoning (DMR) to produce a sentiment intensity prediction $\widehat{y}\in\mathbb{R}$.
The overall inference pipeline is formulated as
\begin{equation}
\setlength{\abovedisplayskip}{5pt}
\setlength{\belowdisplayskip}{5pt}
\widehat{y} = f_{\mathrm{DMR}}\Big(f_{\mathrm{PCS}}\big(f_{\mathrm{SPB}}\big(f_{\mathrm{Enc}}(\mathbf{X}^{\mathrm{v}},\mathbf{X}^{\mathrm{a}},\mathbf{X}^{\mathrm{t}})\big)\big)\Big)
\end{equation}
where $f_{\mathrm{Enc}}$, $f_{\mathrm{SPB}}$, $f_{\mathrm{PCS}}$, and $f_{\mathrm{DMR}}$ denote modality encoding, the sentiment prototype bank, prototype-conditioned modality selection, and the sentiment reasoning process with dynamic modality reweighting, respectively.

\vspace{-0.2cm}

\subsection{Modality Encoding}
\label{sec:encoding}
The raw features of the textual, audio, and visual modalities exhibit fundamentally different representational forms and dimensionalities. 
To enable downstream sentiment prototypes to process information from different modalities in a comparable dimensionality, the encoding stage maps these heterogeneous representations into the same $d$-dimensional space.
Specifically, the textual modality is first processed by a pre-trained BERT encoder to extract contextualized token-level features, which are then projected to $d$ dimensions via a linear layer and further processed by a single-layer Transformer encoder to capture sequence-level temporal dependencies.
Audio and visual features are similarly projected to $d$ dimensions via their respective linear layers and processed by independent single-layer Transformer encoders to model intra-modal temporal patterns.
Formally, this encoding process produces a sequence representation for each modality $m\in\{\mathrm{v},\mathrm{a},\mathrm{t}\}$ as
\begin{equation}
\setlength{\abovedisplayskip}{5pt}
\setlength{\belowdisplayskip}{5pt}
\mathbf{H}^{m} = f_{\mathrm{Enc}}^{m}(\mathbf{X}^{m}), \quad \mathbf{H}^{m} \in \mathbb{R}^{L_m \times d}
\end{equation}
where $L_m$ denotes the sequence length of modality $m$ and $d$ is the unified hidden dimension.
The three modalities employ entirely independent encoder parameters, since different modalities convey sentiment through fundamentally different mechanisms.

\vspace{-0.2cm}

\subsection{Sentiment Prototype Bank}
\label{sec:spb}
A modality sequence usually contains multiple affective cues with internal structure, such as lexical polarity, emphasis, hesitation, facial tension, and cross-modal inconsistency.
If these cues are aggregated indiscriminately, their distinct roles are entangled in a single representation, which makes later comparison and selection less reliable.
SPB addresses this issue by decomposing each modality sequence into a small set of structured affective components before multimodal fusion.

Specifically, SPB introduces $K$ learnable sentiment prototypes that form a shared affective basis across modalities.
Rather than directly summarizing each modality into one global vector, these prototypes probe the modality sequence from $K$ different affective positions and organize the extracted evidence into prototype-aligned slots.
The SPB maintains a learnable prototype matrix $\mathbf{M}\in\mathbb{R}^{K\times d}$, where the $K$ prototype vectors $\{\mathbf{m}_1,\dots,\mathbf{m}_K\}$ gradually specialize during training.
Each prototype is expected to capture one recurrent type of sentiment-relevant evidence, so different slots focus on different affective aspects instead of mixing all cues together.
These prototypes serve as queries that extract structured responses from each modality sequence through cross-attention.
For modality $m$, the prototype-guided decomposition process is defined as
\begin{equation}
\setlength{\abovedisplayskip}{5pt}
\setlength{\belowdisplayskip}{5pt}
\widetilde{\mathbf{Z}}^{m} = \mathrm{LN}\Big(\mathbf{M} + \mathrm{CrossAttn}\big(\mathbf{Q}\!=\!\mathbf{M},\;\mathbf{K}\!=\!\mathbf{H}^{m},\;\mathbf{V}\!=\!\mathbf{H}^{m}\big)\Big)
\end{equation}
\begin{equation}
\setlength{\abovedisplayskip}{5pt}
\setlength{\belowdisplayskip}{5pt}
\mathbf{Z}^{m} = \mathrm{LN}\Big(\widetilde{\mathbf{Z}}^{m} + \mathrm{FFN}(\widetilde{\mathbf{Z}}^{m})\Big)
\end{equation}
where $\mathrm{LN}(\cdot)$ denotes layer normalization, $\mathrm{FFN}(\cdot)$ is a feed-forward network, and $\mathbf{Z}^{m}\in\mathbb{R}^{K\times d}$ is the prototype-aligned response matrix for modality $m$.
Its $k$-th row $\mathbf{z}^{m}_{k}\in\mathbb{R}^{d}$ represents the response of modality $m$ to the $k$-th sentiment prototype.

All three modalities share the same prototype matrix $\mathbf{M}$ but use independent cross-attention parameters.
This design preserves modality-specific extraction while keeping slot semantics consistent across modalities.
As a result, the $k$-th row of text, audio, and visual responses corresponds to the same prototype query, even though the attended temporal positions may differ across modalities.
SPB therefore produces two outcomes at the same time.
First, it compresses each variable-length modality sequence into a fixed-length set of prototype-level responses without collapsing all affective evidence into one undifferentiated vector.
Second, it establishes slot-wise structural correspondence across modalities under a shared affective basis, so responses at the same slot become directly comparable across modalities.
This shared slot structure is critical for the later stages of PRISM.
Because each modality is decomposed with respect to the same set of affective queries, subsequent modality evaluation no longer relies on isolated global summaries, but on aligned prototype-level evidence.
This gives the model a consistent basis for cross-modal comparison, modality reliability estimation, and adaptive fusion.

\vspace{-0.2cm}
\subsection{Prototype-Conditioned Modality Selection}
\label{sec:selection}

After the SPB decomposes the three modalities into slot-corresponding prototype responses, modality evaluation can be performed on aligned affective evidence rather than isolated global summaries.
This raises a key question: for a given sample, how reliable is each modality for sentiment inference at each prototype slot?
Prototype-conditioned modality selection is designed to answer this question.
Its central idea is to assess modality reliability under the same shared affective basis established by SPB, so that modality comparison is carried out separately for each prototype-conditioned affective component instead of once on a collapsed multimodal representation.

Concretely, the reliability of modality $m$ at the $k$-th prototype slot is evaluated by a scoring function.
The score is conditioned on both the modality-specific prototype response $\mathbf{z}^{m}_{k}$ and the corresponding shared prototype $\mathbf{m}_{k}$.
In this way, the model does not estimate a generic modality confidence from a global summary, but judges how reliable modality $m$ is with respect to the specific affective aspect queried by the $k$-th prototype.
The reliability score of modality $m$ at the $k$-th prototype is defined as
\begin{equation}
\setlength{\abovedisplayskip}{5pt}
\setlength{\belowdisplayskip}{5pt}
e_{k}^{m} = g\left([\mathbf{z}^{m}_{k};\;\mathbf{m}_{k}]\right)
\end{equation}
where $g(\cdot)$ is a multi-layer perceptron (MLP) with ReLU activation and $[\cdot;\cdot]$ denotes vector concatenation.
Because the same prototype $\mathbf{m}_{k}$ is shared across modalities, the resulting scores are computed with respect to a consistent reference basis, which makes slot-wise modality comparison well defined.
At each prototype slot, the reliability scores are normalized across modalities into weights and used to perform weighted fusion of the three modality-specific responses.
The normalization and fusion are defined as
\begin{equation}
\setlength{\abovedisplayskip}{5pt}
\setlength{\belowdisplayskip}{5pt}
\alpha_{k}^{m} = \frac{\exp(e_{k}^{m})}{\sum\limits_{m'\in\{\mathrm{v},\mathrm{a},\mathrm{t}\}} \exp(e_{k}^{m'})}
\end{equation}
\begin{equation}
\setlength{\abovedisplayskip}{5pt}
\setlength{\belowdisplayskip}{5pt}
\mathbf{z}_{k} = \sum_{m\in\{\mathrm{v},\mathrm{a},\mathrm{t}\}} \alpha_{k}^{m}\,\mathbf{z}^{m}_{k}
\end{equation}
where $\exp(\cdot)$ denotes the exponential function, $\alpha_{k}^{m}\in[0,1]$ is the normalized weight of modality $m$ at the $k$-th prototype slot and satisfies $\sum_{m}\alpha_{k}^{m}=1$, and $\mathbf{z}_{k}$ is the fused prototype-level token at the $k$-th slot. By assembling all $K$ fused prototype tokens, we obtain
\begin{equation}
\setlength{\abovedisplayskip}{5pt}
\setlength{\belowdisplayskip}{5pt}
\mathbf{Z}_{\mathrm{fused}}=[\mathbf{z}_{1},\dots,\mathbf{z}_{K}]^{\top}\in\mathbb{R}^{K\times d}
\end{equation}
which serves as the structured multimodal representation for subsequent reasoning.
Importantly, this fusion process preserves the prototype-level organization established by the SPB.
Instead of collapsing multimodal evidence into a single undifferentiated vector, it produces a fused representation in which each row still corresponds to a specific prototype-conditioned affective component.
This enables later reasoning stages to operate on structured multimodal evidence while retaining explicit modality evaluation at each slot.

\subsection{Dynamic Modality Reweighting}
\label{sec:dmr}

The previous prototype-conditioned selection stage aggregates multimodal evidence at each slot and produces fused prototype tokens $\mathbf{Z}_{\mathrm{fused}}$.
However, the modality importance estimated at fusion time does not necessarily remain optimal throughout subsequent reasoning.
As the representation evolves across layers, the model may need to re-assess whether a modality still provides reliable evidence under the current context.
Therefore, we retain the modality-specific prototype responses and regulate their influence throughout Transformer backbone reasoning.

The input to the Transformer backbone is a concatenation of five groups of tokens.
It consists of a learnable classification token $\mathbf{h}_{\mathrm{cls}}\in\mathbb{R}^{d}$, followed by the $K$ fused prototype tokens $\mathbf{Z}_{\mathrm{fused}}$, and the $K$ modality-specific prototype responses from text, audio, and visual modalities, namely $\mathbf{Z}^{\mathrm{t}}$, $\mathbf{Z}^{\mathrm{a}}$, and $\mathbf{Z}^{\mathrm{v}}$.
The total sequence length is $1+4K$, and learnable positional embeddings are added to all tokens.
This token organization allows the model to reason over integrated evidence while preserving direct access to modality-specific prototype evidence.
The backbone contains $L$ Transformer layers.
At the $l$-th layer, the whole sequence first passes through a pre-norm self-attention block and a pre-norm feed-forward block for global interaction.
Let $\mathbf{h}_{\mathrm{cls}}^{(l)}$ denote the updated classification token after these two operations.
Since this token summarizes the current global context, we use it to generate a gate vector that determines how strongly each modality should be preserved at this layer:
\begin{equation}
\setlength{\abovedisplayskip}{5pt}
\setlength{\belowdisplayskip}{5pt}
\mathbf{g}^{(l)} = \sigma\!\left(\mathbf{W}_{l}\,\mathbf{h}_{\mathrm{cls}}^{(l)} + \mathbf{b}_{l}\right) \in \mathbb{R}^{3}
\end{equation}
where $\mathbf{W}_{l}$ and $\mathbf{b}_{l}$ are learnable parameters of the $l$-th layer, and $\sigma(\cdot)$ is the sigmoid function.
The three elements of $\mathbf{g}^{(l)}$ correspond to the textual, audio, and visual modalities.
For each modality, the corresponding gate value uniformly rescales all its $K$ prototype tokens:
\begin{equation}
\setlength{\abovedisplayskip}{5pt}
\setlength{\belowdisplayskip}{5pt}
\mathbf{Z}^{m,(l)} \leftarrow g_{m}^{(l)} \cdot \mathbf{Z}^{m,(l)}, \quad m \in \{\mathrm{t},\mathrm{a},\mathrm{v}\}
\end{equation}
Notably, the classification token $\mathbf{h}_{\mathrm{cls}}$ and the fused prototype tokens $\mathbf{Z}_{\mathrm{fused}}$ are not gated.
After $L$ layers of reasoning, the final hidden state of the classification token is fed into a regression head to predict sentiment intensity.
The regression head is implemented as a MLP that outputs the prediction $\widehat{y}$.
In this way, modality contribution is not only estimated during prototype-conditioned fusion, but also revised layer by layer throughout subsequent reasoning.

\subsection{Prototype Learning Objectives}
\label{sec:training}
PRISM is optimized by three objectives for sentiment prediction, prototype informativeness, and prototype diversity.
The primary objective is the mean squared error loss:
\begin{equation}
\setlength{\abovedisplayskip}{5pt}
\setlength{\belowdisplayskip}{5pt}
\mathcal{L}_{\mathrm{reg}} = \mathbb{E}\!\left[\left(\widehat{y} - y\right)^{2}\right]
\end{equation}
where $\mathbb{E}[\cdot]$ denotes the expectation over training samples, and $y$ is the ground-truth sentiment intensity.
To make each fused prototype representation in $\mathbf{Z}_{\mathrm{fused}}$ sentiment-discriminative, we further impose auxiliary supervision on every prototype slot. Specifically, each fused prototype representation $\mathbf{z}_k$ in $\mathbf{Z}_{\mathrm{fused}}$ is fed into a lightweight linear head to predict the same utterance-level label:
\begin{equation}
\setlength{\abovedisplayskip}{5pt}
\setlength{\belowdisplayskip}{5pt}
\mathcal{L}_{\mathrm{aux}} = \mathbb{E}\!\left[\frac{1}{K}\sum_{k=1}^{K}\left(\widehat{y}_{k}^{\mathrm{aux}} - y\right)^{2}\right]
\end{equation}
where $\widehat{y}_{k}^{\mathrm{aux}}$ is the auxiliary prediction from the $k$-th fused prototype representation $\mathbf{z}_k$. This objective encourages each prototype to retain useful sentiment information, while distinct prototype-conditioned cross-attention paths and the diversity term prevent trivial collapse.
We also introduce a diversity regularization term to encourage complementary prototypes. Let $\overline{\mathbf{M}} \in \mathbb{R}^{K \times d}$ denote the row-wise $\ell_2$-normalized prototype matrix $\mathbf{M}$, where $\overline{\mathbf{m}}_k = \mathbf{m}_k / \|\mathbf{m}_k\|_2$. The diversity loss is
\begin{equation}
\setlength{\abovedisplayskip}{5pt}
\setlength{\belowdisplayskip}{5pt}
\mathcal{L}_{\mathrm{div}} = \left\|\overline{\mathbf{M}}\,\overline{\mathbf{M}}^{\top} - \mathbf{I}\right\|_{\mathrm{F}}^{2}
\end{equation}
where $\|\cdot\|_{\mathrm{F}}$ denotes the Frobenius norm. Minimizing $\mathcal{L}_{\mathrm{div}}$ encourages different prototypes to remain separated and capture diverse affective patterns.
The overall objective is
\begin{equation}
\setlength{\abovedisplayskip}{5pt}
\setlength{\belowdisplayskip}{5pt}
\mathcal{L} = \mathcal{L}_{\mathrm{reg}} + \lambda_{\mathrm{aux}}\,\mathcal{L}_{\mathrm{aux}} + \lambda_{\mathrm{div}}\,\mathcal{L}_{\mathrm{div}}
\end{equation}
where $\lambda_{\mathrm{aux}}$ and $\lambda_{\mathrm{div}}$ control the contributions of the auxiliary and diversity terms.
\begin{table}[t]
\centering
\caption{Statistics of the CMU-MOSI, CMU-MOSEI, and CH-SIMS datasets, including sentiment score ranges and the numbers of training, validation, test, and total samples.}
\label{tab:datasets}
\vspace{-0.3cm}
\begin{tabular}{lcrrrr}
\toprule
\textbf{Dataset} & \textbf{Score Range} & \textbf{Train} & \textbf{Valid} & \textbf{Test} & \textbf{Total}  \\
\midrule
CMU-MOSI & $[-3, +3]$ & 1,284  & 229   & 686   & 2,199   \\
CMU-MOSEI & $[-3, +3]$ & 16,326 & 1,871 & 4,659 & 22,856  \\
CH-SIMS   & $[-1, +1]$ & 1,368  & 456   & 457   & 2,281   \\
\bottomrule
\end{tabular}
\vspace{-0.5cm}
\end{table}

\begin{table*}[t]
\centering
\caption{Ablation study on CMU-MOSI, CMU-MOSEI, and CH-SIMS. Each row removes one component from the full PRISM model. For CMU-MOSI and CMU-MOSEI, Acc-2 and F1 are reported in NN/NP format (\%).}
\label{tab:ablation}
\vspace{-0.3cm}
\resizebox{\linewidth}{!}{
\begin{tabular}{l|ccccc|ccccc|ccccc}
\toprule
\multirow{2}{*}{Variant} & \multicolumn{5}{c|}{CMU-MOSI} & \multicolumn{5}{c|}{CMU-MOSEI} & \multicolumn{5}{c}{CH-SIMS} \\
\cmidrule(lr){2-6}\cmidrule(lr){7-11}\cmidrule(lr){12-16}
& MAE$\downarrow$ & Corr$\uparrow$ & Acc-7$\uparrow$ & Acc-2$\uparrow$ & F1$\uparrow$ & MAE$\downarrow$ & Corr$\uparrow$ & Acc-7$\uparrow$ & Acc-2$\uparrow$ & F1$\uparrow$ & MAE$\downarrow$ & Corr$\uparrow$ & Acc-3$\uparrow$ & Acc-2$\uparrow$ & F1$\uparrow$ \\
\midrule
PRISM (Full)         & \textbf{0.691} & \textbf{0.813} & \textbf{47.25} & \textbf{84.64}/\textbf{86.61} & \textbf{84.58}/\textbf{86.56} & \textbf{0.519} & \textbf{0.783} & \textbf{54.55} & \textbf{84.41}/\textbf{86.57} & \textbf{84.83}/\textbf{86.68} & \textbf{0.405} & \textbf{0.617} & \textbf{66.88} & \textbf{81.24} & \textbf{81.28} \\
\midrule
w/o SPB                & 0.756 & 0.777 & 45.19 & 80.78/82.23 & 80.74/82.25 & 0.538 & 0.761 & 52.34 & 82.52/84.38 & 82.65/84.27 & 0.434 & 0.522 & 63.61 & 76.82 & 77.05 \\
w/o Selection & 0.757 & 0.779 & 45.34 & 80.76/82.16 & 80.79/82.25 & 0.529 & 0.773 & 53.45 & 83.25/85.45 & 83.60/85.15 & 0.415 & 0.600 & 65.72 & 79.87 & 79.85 \\
w/o Fine Path          & 0.754 & 0.771 & 45.63 & 81.78/83.23 & 81.71/83.21 & 0.522 & 0.778 & 54.09 & 84.12/86.08 & 84.35/85.98 & 0.409 & 0.596 & 65.28 & 77.90 & 78.23 \\
w/o DMR Gates          & 0.732 & 0.777 & 45.48 & 81.92/83.38 & 81.88/83.39 & 0.534 & 0.767 & 52.18 & 82.19/84.29 & 82.63/84.68 & 0.427 & 0.548 & 64.28 & 77.43 & 78.33 \\
w/o Shared Proto       & 0.734 & 0.785 & 45.04 & 83.19/84.71 & 83.14/84.71 & 0.525 & 0.775 & 54.07 & 82.72/84.85 & 82.95/84.73 & 0.415 & 0.599 & 66.01 & 78.77 & 78.99 \\
\bottomrule
\end{tabular}
}
\end{table*}

\begin{table*}[t]
\centering
\caption{Comparison with representative MSA approaches on CMU-MOSI, CMU-MOSEI, and CH-SIMS. Best results in each column are \textbf{bolded} while second-best are \underline{underlined}. The specific meaning of each symbol is as follows: ``$^{\dagger}$'': Results from \citet{mao2022mmsa}. ``$^*$'': Results reproduced from code provided by their authors.``$^\ddagger$'': Results from original paper. ``--'': Not provided in original paper. For CMU-MOSI and CMU-MOSEI, Acc-2 and F1 are reported in NN/NP format (\%).}
\vspace{-0.3cm}
\label{tab:main_results}
\resizebox{\linewidth}{!}{
\begin{tabular}{l|ccccc|ccccc|ccccc}
\toprule
\multirow{2}{*}{Methods} & \multicolumn{5}{c|}{CMU-MOSI} & \multicolumn{5}{c|}{CMU-MOSEI} & \multicolumn{5}{c}{CH-SIMS} \\
\cmidrule(lr){2-6}\cmidrule(lr){7-11}\cmidrule(lr){12-16}
& MAE$\downarrow$ & Corr$\uparrow$ & Acc-7$\uparrow$ & Acc-2$\uparrow$ & F1$\uparrow$ & MAE$\downarrow$ & Corr$\uparrow$ & Acc-7$\uparrow$ & Acc-2$\uparrow$ & F1$\uparrow$ & MAE$\downarrow$ & Corr$\uparrow$ & Acc-3$\uparrow$ & Acc-2$\uparrow$ & F1$\uparrow$ \\
\midrule
LMF$^{\dagger}$      & 0.950 & 0.651 & 33.82 & 77.90/79.18 & 77.80/79.15 & 0.576 & 0.717 & 51.59 & 80.54/83.48 & 80.94/83.36 & 0.441 & 0.576 & 64.68 & 77.77 & 77.88 \\
MulT$^{\dagger}$     & 0.879 & 0.702 & 36.91 & 79.71/80.98 & 79.63/80.95 & 0.559 & 0.733 & 52.84 & 81.15/84.63 & 81.56/84.52 & 0.453 & 0.564 & 64.77 & 78.56 & 79.66 \\
MISA$^{\dagger}$     & 0.776 & 0.778 & 41.37 & 81.84/83.54 & 81.82/83.58 & 0.557 & 0.751 & 52.05 & 80.67/84.67 & 81.12/84.66 & -- & -- & -- & -- & -- \\
ConFEDE$^*$          & 0.742 & 0.784 & 42.27 & 84.17/85.52 & 84.13/85.52 & 0.532 & 0.768 & 53.16 & 81.65/85.82 & 82.17/85.83 & -- & -- & -- & -- & -- \\
ALMT$^{\dagger}$  & 0.712 & 0.793 & 46.79 & 83.97/85.82 & 84.05/85.86 & 0.530 & 0.774 & 53.62 & 81.54/85.99 & 81.05/86.05 & 0.408 & 0.594 & 65.86 & 78.77 & 78.71 \\
DBF$^\ddagger$      & \underline{0.693} & 0.801 & 44.80 & \textbf{85.10}/\textbf{86.90} & \textbf{85.10}/\textbf{86.90} & \underline{0.523} & 0.772 & \underline{54.20} & \underline{84.30}/86.40 & \underline{84.80}/86.20 & -- & -- & -- & -- & -- \\
KuDA$^{\dagger}$     & 0.705 & 0.795 & \underline{47.08} & 84.40/86.43 & 84.48/86.46 & 0.529 & 0.776 & 52.89 & 83.26/\underline{86.46} & 82.97/\underline{86.59} & 0.408 & 0.613 & 66.52 & 80.74 & 80.71 \\
SIMSUF$^\ddagger$   & 0.709 & \underline{0.802} & 45.72 & --/86.08     & --/85.98     & 0.529 & 0.772 & 53.68 & --/86.23     & --/86.12     & -- & -- & -- & -- & -- \\
DashFusion$^*$           & 0.816 & 0.766 & 40.47 & 82.30/83.84 & 82.22/83.82 & 0.528 & \underline{0.779} & 53.34 & 81.05/85.65 & 81.66/85.69 & 0.414 & 0.599 & 66.68 & 77.84 & 78.24 \\
DPDF-LQ$^*$              & 0.742 & 0.783 & 44.64 & 83.24/85.61 & 83.03/85.50 & 0.562 & 0.765 & 50.90 & 82.39/85.83 & 82.79/85.76 & 0.416 & 0.595 & 65.21 & 78.34 & 78.44 \\
\midrule
PRISM (Ours)         & \textbf{0.691} & \textbf{0.813} & \textbf{47.25} & \underline{84.64}/\underline{86.61} & \underline{84.58}/\underline{86.56} & \textbf{0.519} & \textbf{0.783} & \textbf{54.55} & \textbf{84.41}/\textbf{86.57} & \textbf{84.83}/\textbf{86.68} & \textbf{0.405} & \textbf{0.617} & \textbf{66.88} & \textbf{81.24} & \textbf{81.28} \\
\bottomrule
\end{tabular}
}
\end{table*}

\section{Experiment Settings}
\subsection{Datasets}
\label{sec:datasets}

We evaluate PRISM on three widely used multimodal sentiment analysis benchmarks, whose statistics are summarized in Table~\ref{tab:datasets}. 
CMU-MOSI~\cite{zadeh2016multimodal} and CMU-MOSEI~\cite{zadeh2018multi} are English video sentiment datasets, while CH-SIMS~\cite{yu2020ch} is a Chinese dataset. 
CMU-MOSI provides a relatively small benchmark, CMU-MOSEI offers a larger and more diverse testbed, and CH-SIMS extends evaluation to a different language and label granularity.

Following prior work \cite{yu2021learning, liang2021attention, lv2021progressive}, we report mean absolute error (MAE) and Pearson correlation (Corr) for regression evaluation. 
For CMU-MOSI and CMU-MOSEI, we further report 7-class accuracy (Acc-7), as well as binary accuracy (Acc-2) and F1 score in two forms: negative/non-negative (NN, including zero-labeled samples) and negative/positive (NP, excluding zero-labeled samples).
For CH-SIMS, we report Acc-2, three-class accuracy (Acc-3), and F1.

\subsection{Baselines}
\label{sec:baselines}

We evaluate PRISM through both ablation studies and comparisons with representative prior methods.

For ablation, we construct five variants to assess the contribution of each core component: 
``w/o SPB'' replaces shared-prototype extraction with mean pooling; 
``w/o Selection'' uses uniform modality weights; 
``w/o Fine Path'' removes the modality-specific fine-grained response paths and keeps only $\mathbf{Z}_{\mathrm{fused}}$ as backbone input; 
``w/o DMR Gates'' removes the layer-wise sigmoid gating mechanism; 
and ``w/o Shared Proto'' replaces the shared prototype bank with modality-specific prototype matrices.

For comparison with prior work, we include $10$ representative MSA methods spanning early fusion, cross-modal attention, representation decomposition, text-guided fusion, dynamic modality weighting, and bottleneck-based fusion: LMF~\cite{liu2018efficient}, MulT~\cite{tsai2019multimodal}, MISA~\cite{hazarika2020misa}, ConFEDE~\cite{yang2023confede}, ALMT~\cite{zhang2023almt}, DBF~\cite{wu2023dbm}, KuDA~\cite{feng2024kuda}, SIMSUF~\cite{huang2024simsuf}, DashFusion~\cite{wen2025dashfusion}, and DPDF-LQ~\cite{zhou2025dpdf}.

\subsection{Implementation Details}
\label{sec:implementation}

Following previous work \cite{zhang2023almt, zhou2025dpdf, wen2025dashfusion}, we use pre-extracted feature sequences for all modalities, where textual, visual, and acoustic features are obtained from BERT \cite{devlin2019bert}, OpenFace \cite{zadeh2018openface}, and LibROSA \cite{mcfee2015librosa}, respectively. Across all datasets, we set the number of sentiment prototypes to $K{=}8$, the hidden dimension to 128, the number of attention heads to 8. We use a 2-layer Transformer backbone on CMU-MOSI and a 4-layer one on CMU-MOSEI and CH-SIMS.
We train PRISM with AdamW \cite{loshchilov2017decoupled} using differential learning rates for BERT and the remaining parameters, with linear warmup followed by cosine annealing \cite{loshchilov2016sgdr}. 
The batch size is 64 and dropout is 0.1. 
Results are averaged over five random seeds. 
All experiments are conducted on an NVIDIA RTX 5090 GPU.

\begin{figure*}[t]
\centering
\includegraphics[width=1\linewidth, trim=0 20 0 0]{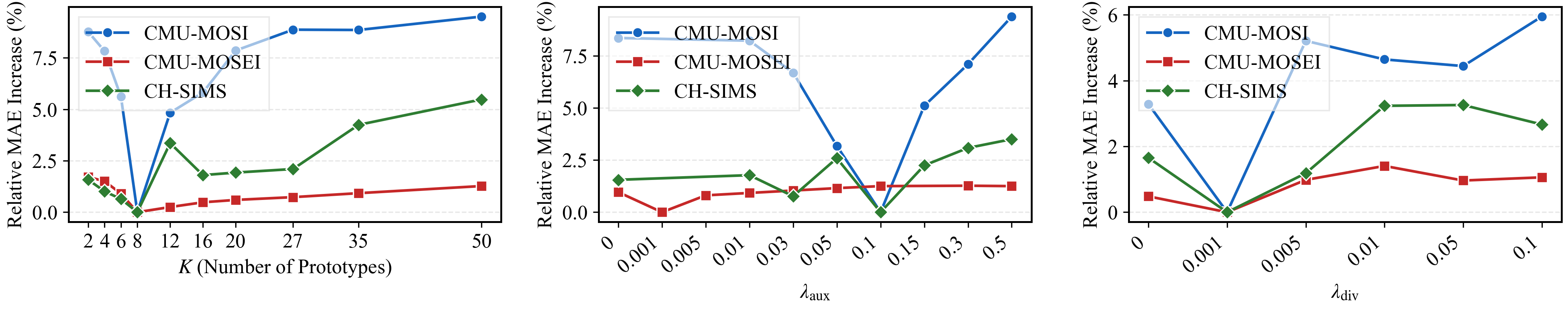}
\caption{Hyperparameter sensitivity on CMU-MOSI (blue), CMU-MOSEI (red), and CH-SIMS (green). Left: number of sentiment prototypes $K$. Middle: auxiliary loss weight $\lambda_{\text{aux}}$. Right: diversity loss weight $\lambda_{\text{div}}$. All curves report relative MAE increase (\%) from each dataset's optimal value, enabling cross-dataset comparison despite different absolute MAE ranges.}
\label{fig:sensitivity}
\vspace{-0.2cm}
\end{figure*}

\begin{table*}
\centering
\caption{Effect of missing modalities on CMU-MOSI, CMU-MOSEI, and CH-SIMS. Missing modalities are replaced with zero vectors. For CMU-MOSI and CMU-MOSEI, Acc-2 is reported in NP format (\%).}
\label{tab:modality_robust}
\vspace{-0.2cm}
\resizebox{\linewidth}{!}{
\begin{tabular}{l|cccc|cccc|cccc}
\toprule
\multirow{2}{*}{Condition} & \multicolumn{4}{c|}{CMU-MOSI} & \multicolumn{4}{c|}{CMU-MOSEI} & \multicolumn{4}{c}{CH-SIMS} \\
\cmidrule(lr){2-5}\cmidrule(lr){6-9}\cmidrule(lr){10-13}
& MAE$\downarrow$ & Corr$\uparrow$ & Acc-7$\uparrow$ & Acc-2$\uparrow$ & MAE$\downarrow$ & Corr$\uparrow$ & Acc-7$\uparrow$ & Acc-2$\uparrow$ & MAE$\downarrow$ & Corr$\uparrow$ & Acc-3$\uparrow$ & Acc-2$\uparrow$ \\
\midrule
Full     & \textbf{0.691} & \textbf{0.813} & \textbf{47.25} & \textbf{86.61} & \textbf{0.519} & \textbf{0.783} & \textbf{54.55} & \textbf{86.57} & \textbf{0.405} & \textbf{0.617} & \textbf{66.88} & \textbf{81.24} \\
\midrule
w/o V    & 0.723 & 0.753 & 46.10 & 84.71 & 0.534 & 0.769 & 53.68 & 84.48 & 0.436 & 0.577 & 64.33 & 74.62 \\
w/o A    & 0.723 & 0.756 & 46.15 & 84.67 & 0.542 & 0.775 & 53.40 & 85.28 & 0.422 & 0.597 & 63.68 & 78.77 \\
w/o T    & 1.352 & 0.121 & 18.66 & 57.46 & 1.007 & 0.053 & 38.36 & 62.85 & 0.570 & 0.342 & 54.05 & 70.24 \\
T only   & 0.725 & 0.752 & 46.26 & 84.37 & 0.540 & 0.768 & 53.55 & 85.09 & 0.441 & 0.572 & 63.46 & 74.84 \\
A only   & 1.356 & 0.107 & 18.51 & 57.47 & 0.993 & 0.072 & 38.08 & 62.83 & 0.587 & 0.085 & 48.14 & 69.37 \\
V only   & 1.355 & 0.111 & 18.80 & 57.43 & 1.069 & 0.059 & 38.21 & 62.87 & 0.577 & 0.300 & 51.64 & 69.80 \\
\bottomrule
\end{tabular}
}
\vspace{-0.1cm}
\end{table*}

\begin{figure*}[t]
\centering
\includegraphics[width=1\linewidth, trim=0 20 0 0]{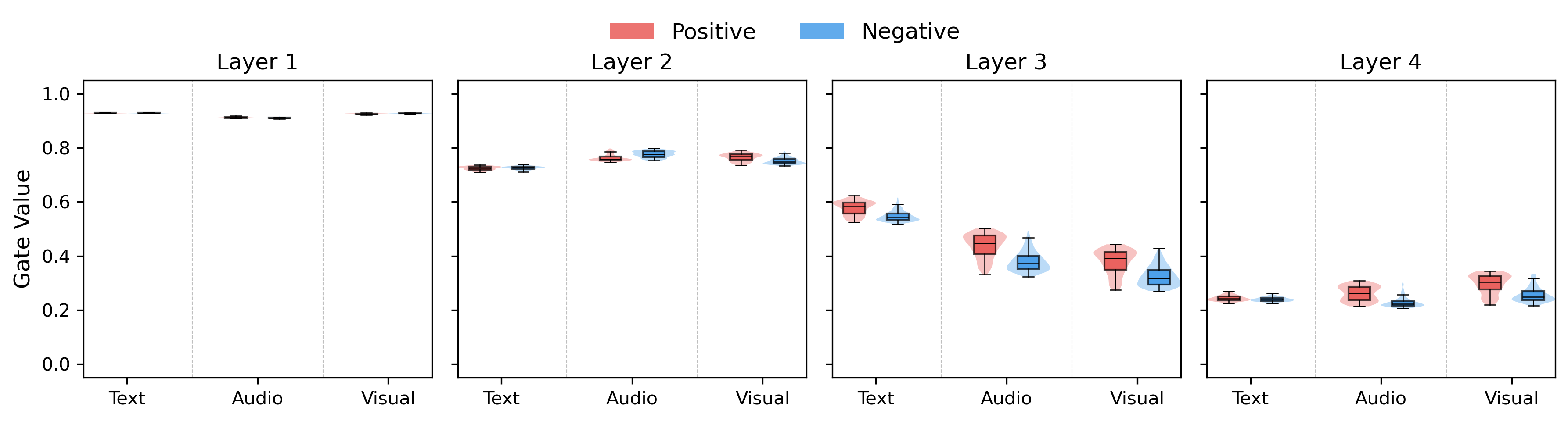}
\caption{Layer-wise DMR gate distributions on the CH-SIMS test set. Violin and box plots show gate values for text, audio, and visual modalities in positive and negative samples across all Transformer backbone layers.}
\label{fig:dmr_gates}
\end{figure*}

\section{Results and Analysis}
\subsection{Overall Results}
Table~\ref{tab:ablation} reports the results of five ablation variants on all three datasets. All variants underperform the full model, which shows that each component contributes to the final prediction quality. Replacing the SPB with mean pooling consistently degrades performance, indicating that simple coarse compression cannot recover the diverse affective cues distributed across modalities and that structured multi-prototype extraction is necessary. Fixing the modality weights to equal values also hurts performance, which confirms that sentiment prediction benefits from sample-dependent and prototype-dependent modality selection rather than static fusion.
Removing the fine path further reduces performance, showing that fused tokens alone are not sufficient and that modality-specific evidence remains useful after selection. Removing the DMR gates also leads to clear degradation, which suggests that the model benefits from layer-wise control over how fine-grained modality evidence is injected into reasoning. Replacing shared prototypes with independent modality-specific prototypes weakens performance as well, supporting the use of a shared prototype bank as a common basis for cross-modal comparison and reliability estimation.

Table \ref{tab:main_results} compares the PRISM with representative MSA approaches across several fusion paradigms on all three datasets.
Early symmetric fusion methods such as LMF and MulT remain the weakest overall. Because they treat all modalities uniformly, noisy or weak cues from less informative modalities are mixed with useful evidence, which limits sentiment modeling on both English benchmarks and CH-SIMS.
Methods that explicitly model modality asymmetry perform better. Decomposition-based methods such as MISA and ConFEDE separate shared and private information, while text-centered methods such as ALMT exploit the dominant role of language. However, these approaches still rely on either sample-agnostic decomposition or a fixed modality hierarchy, which limits their ability to handle varying modality reliability across samples.
Stronger baselines such as DBF, KuDA, DashFusion, and DPDF-LQ further improve fusion through bottleneck compression, dynamic weighting, or refined interaction modeling, but the PRISM still achieves the best overall performance. Specifically, on CMU-MOSI, PRISM reduces MAE by 0.014 and improves Corr by 0.018 over KuDA. On CMU-MOSEI, it improves Acc-7 by 1.21 points over DashFusion while also achieving the best MAE and Corr. On CH-SIMS, the PRISM outperforms the best competing results on all five metrics, including gains of 0.50 points in Acc-2 and 0.57 points in F1. These results show that the shared-prototype design provides a stronger basis for cross-modal comparison and adaptive modality integration than existing fusion strategies.

\vspace{-0.3cm}

\subsection{Hyperparameter Sensitivity}
\label{sec:sensitivity}
To examine the robustness of PRISM, we vary three key hyperparameters, and report the results in Figure \ref{fig:sensitivity}.

\textbf{Effect of prototype number $K$.}
Figure \ref{fig:sensitivity} (left) shows that all three datasets perform best at $K{=}8$. A small $K$ makes the prototype space too coarse to capture diverse affective patterns, while a large $K$ introduces redundancy and weakens representation quality. The consistent optimum across datasets suggests that a moderate prototype granularity is sufficient.

\textbf{Effect of auxiliary loss weight $\lambda_{\text{aux}}$.}
As shown in Figure \ref{fig:sensitivity} (middle), $\mathcal{L}_{\text{aux}}$ is more important on CMU-MOSI and CH-SIMS than on CMU-MOSEI. On the two smaller datasets, removing or underweighting it causes clear degradation, and $\lambda_{\text{aux}}{=}0.1$ gives the best result. By contrast, CMU-MOSEI remains relatively stable, which suggests that larger-scale data can better organize the prototype space from the main regression signal alone.

\textbf{Effect of diversity regularization weight $\lambda_{\text{div}}$.}
Figure \ref{fig:sensitivity} (right) shows that a small positive value, $\lambda_{\text{div}}{=}0.001$, gives the best or near-best result on all three datasets. Without this term, prototype collapse is more likely; with overly large values, excessive separation harms performance. This suggests that prototype diversity is beneficial, but only under mild regularization.

\begin{figure*}[t]
\centering
\includegraphics[width=1\linewidth, trim=0 12 0 0]{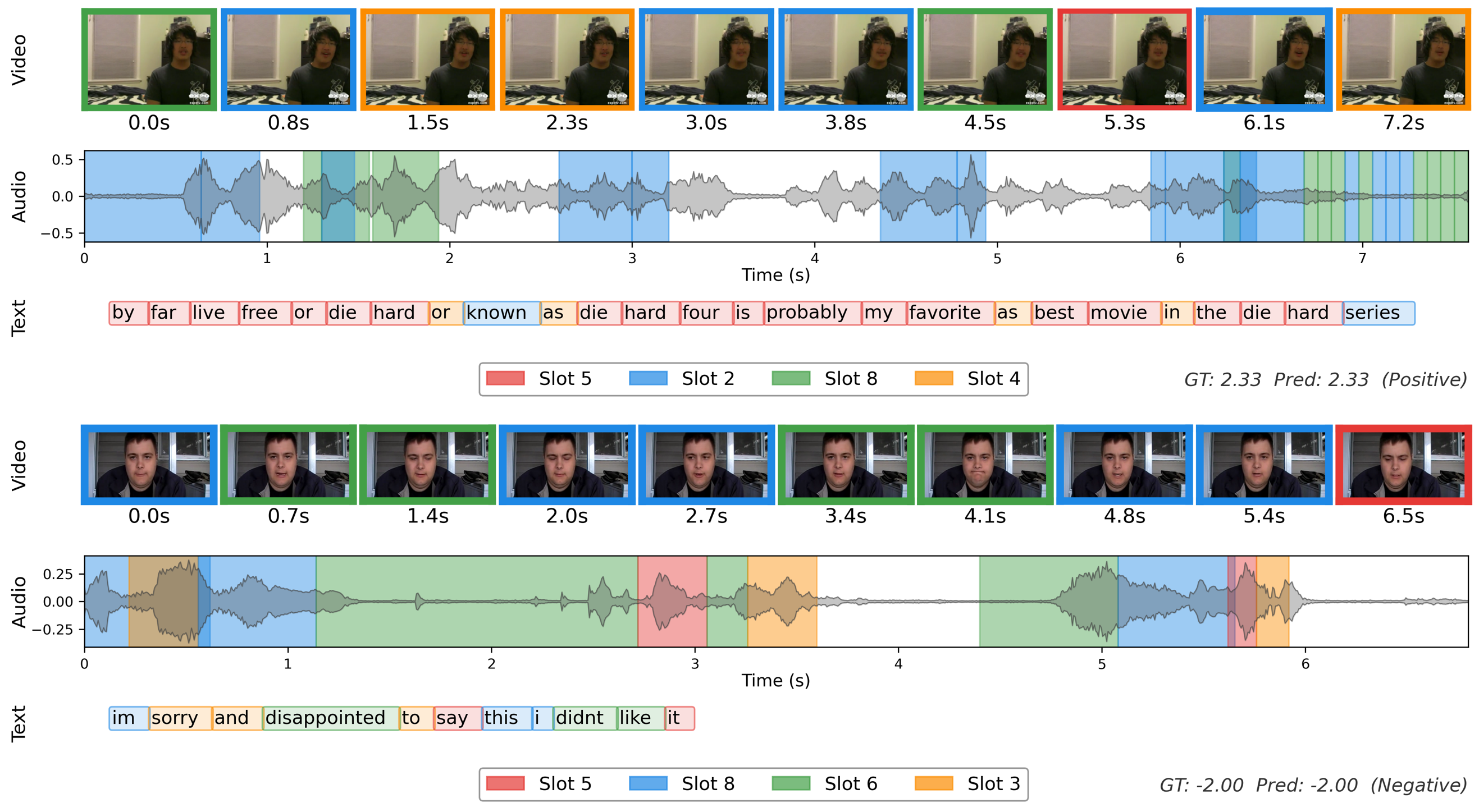}
\caption{Visualization of sentiment prototypes for two samples where each slot corresponds to one sentiment prototype. Each panel shows, from top to bottom: video frames with prototype-colored borders, audio waveform with temporal prototype overlay and text with word-level prototype highlighting.}
\label{fig:case_study}
\vspace{-0.3cm}
\end{figure*}

\vspace{-0.2cm}
\subsection{Effect of Modalities}
\label{sec:analysis_robust}
To examine whether PRISM learns calibrated modality reliance, we evaluate it under six missing-modality conditions by masking one or two modalities. Results are reported in Table~\ref{tab:modality_robust}.
Across all three datasets, removing text causes the largest degradation, confirming that text remains the primary carrier of explicit sentiment cues. This shows that PRISM does not enforce balanced fusion, but relies most on the modality with the strongest sentiment signal.
The role of non-textual modalities differs across datasets. On CMU-MOSI and CMU-MOSEI, the text-only setting remains close to the full model, suggesting that audio and visual modalities mainly provide supplementary cues. In contrast, CH-SIMS shows a larger multimodal gain: MAE increases from 0.405 to 0.441 in the text-only setting, and removing vision alone already degrades MAE to 0.436. This indicates that non-textual signals contribute more independently on CH-SIMS, and PRISM can exploit them more effectively.
When only one non-textual modality is retained, performance drops sharply on all three datasets, showing that audio or visual cues alone are insufficient for reliable prediction. On CH-SIMS, visual-only is slightly stronger than audio-only, suggesting greater independent discriminative value from vision. Overall, PRISM adapts its modality reliance to the informativeness structure of each dataset: it remains strongly text-dominant on CMU-MOSI and CMU-MOSEI, while making fuller use of non-textual evidence on CH-SIMS.

\vspace{-0.2cm}

\subsection{Effect of Layer-wise DMR}
\label{sec:analysis_weights}

To examine how the DMR module regulates modality evidence during reasoning, we visualize its gate distributions for positive and negative samples on CH-SIMS across backbone layers in Figure \ref{fig:dmr_gates}. A clear depth-dependent pattern emerges: DMR is nearly transparent in shallow layers and becomes selective only in deeper layers.
In Layers 1 and 2, all three modalities retain high gate values with only minor polarity differences, indicating that these layers mainly preserve fused multimodal evidence rather than strongly reweight it. The main effect of DMR appears in Layer 3, where gate values drop substantially and the separation between positive and negative samples becomes most pronounced. This shows that modality regulation is activated primarily after the representation becomes sufficiently structured, rather than being applied uniformly throughout the backbone.
Layer 4 continues this trend with lower overall gate values, indicating a final stage of selective compression.
These results show that DMR acts as a depth-selective post-fusion regulator. The fusion-stage weights determine the initial modality composition, while DMR further adjusts how much modality-specific evidence is retained as reasoning proceeds.

\vspace{-0.3cm}
\subsection{Case Study}
\label{sec:case_study}

Figure \ref{fig:case_study} shows two correctly predicted utterances from the CMU-MOSEI and CMU-MOSI test sets. In both cases, the same slot highlights temporally corresponding evidence across video, audio, and text, suggesting that PRISM organizes multimodal inputs into coherent affective units rather than relying on isolated cues.
In the positive case, PRISM concentrates on strongly opinion-bearing expressions such as ``\textit{by far}'', ``\textit{probably my favorite}'', and ``\textit{best movie}'', while the corresponding slots also activate aligned video frames and audio regions within the same opinion span. This pattern indicates that the model accumulates consistent positive evidence across modalities instead of depending on a single cue. In the negative case, PRISM focuses on explicit negative expressions such as ``\textit{disappointed}'', ``\textit{didn't like}'', and the concluding ``\textit{it}'', with aligned nonverbal evidence concentrated around the same local regions. This shows that the model identifies the sentiment-critical negative judgment and anchors it to synchronized multimodal context.
These cases show that PRISM supports interpretable multimodal reasoning by structuring cross-modal evidence at the slot level and preserving sentiment-critical local patterns.

\vspace{-0.2cm}
\section{Conclusion}

In this paper, we propose PRISM, a multimodal sentiment analysis framework built on shared sentiment prototypes. By combining structured prototype extraction, adaptive modality selection, and dynamic modality reweighting, the PRISM captures multidimensional affective cues while preserving modality-specific evidence throughout the reasoning process.
Experiments on CMU-MOSI, CMU-MOSEI, and CH-SIMS show that PRISM achieves strong performance across datasets. Further analyses show that the shared prototypes provide an effective basis for cross-modal comparison, structured fusion and modality evaluation. Meanwhile, the layer-wise dynamic reweighting mechanism continuously adjusts modality contributions according to the evolving semantic context.

\bibliographystyle{ACM-Reference-Format}
\bibliography{main}

\end{document}